\documentclass[runningheads]{llncs}
\usepackage{wrapfig}
\usepackage{xcolor}
\usepackage{tabularx}
\usepackage{amsmath}
\usepackage{graphicx}
\usepackage{hyperref}
\usepackage{mathtools}
\usepackage{booktabs}
\usepackage{booktabs}
\usepackage{subcaption}
\usepackage{multirow}
\usepackage{pifont} 
\newcommand{\cmark}{\ding{51}}
\newcommand{\xmark}{\ding{55}}
\usepackage{wrapfig}
\usepackage[misc,geometry]{ifsym}
\usepackage{xcolor}
\usepackage{tabularx}
\usepackage{footnote}
\usepackage{amsmath}
\usepackage{graphicx}
\usepackage{hyperref}
\usepackage{mathtools}
\usepackage{soul}
\usepackage[stable]{footmisc}
\usepackage{graphicx}
\usepackage{subcaption}
\usepackage{booktabs}
\usepackage{amssymb}
\usepackage{multirow}
\usepackage{seqsplit}
\usepackage{tikz}
\usetikzlibrary{fit, backgrounds}
\usetikzlibrary{shapes.geometric, arrows.meta, positioning, calc}

\usepackage{lipsum} 

\begin{document}
\title{Robust Interpretation of Historical Documents in Knowledge Graphs Through Query Inference and Execution}
\titlerunning{Interpreting Historical Documents in Knowledge Graphs}
%
\renewcommand{\thefootnote}{\fnsymbol{footnote}}
\author{
  Sebastià Nicolau\footnotemark[2] \and
  Adrià Molina \footnotemark[2]\hspace{0.25em}\href{mailto:amolina@cvc.uab.cat}{\Letter} \and
  Oriol Ramos Terrades \and
  Josep Lladós
}
\footnotetext[2]{Both authors contributed equally to this work.}
\renewcommand{\thefootnote}{\arabic{footnote}}
%
\authorrunning{S. Nicolau, A. Molina, et al.}

\institute{Centre de Visió per Computador \and Universitat Autònoma de Barcelona \\   \email{\{snicolau, amolina, oriolrt, josep\}@cvc.uab.cat}}
\maketitle              
\begin{abstract} 

The emergence of Large Language Models (LLMs) has redefined how users interact with information in digital environments. However, their widespread and often indiscriminate integration has raised significant concerns regarding reliability and trustworthiness issues that are particularly critical when accessing digital libraries and historical archives. How can one leverage the generalization capacity of an LLM without losing the level of accountability required for an archival institution? In this paper, we present an agentic retrieval system designed to deliver more accurate and verifiable access to historical data while preserving much of the flexibility associated with unconstrained LLMs. As a contribution to historical document analysis, we compare traditional Retrieval-Augmented Generation (RAG) with an agentic GraphRAG architecture in their ability to deliver historical information under realistic conditions, including the presence of OCR and transcription errors.

We introduce a semi-symbolic framework that integrates word-spotting techniques for post-OCR correction with a knowledge graph representation that enables the agent to access information through synthesized queries. The interleaved collaboration between word spotting and code generation allows the agent to construct strong retrieval queries that are robust to misinterpretation and hallucination, while still leveraging approximate search when noise and uncertainty, common in historical document analysis, would otherwise hinder precise retrieval.

\end{abstract}

\section{Introduction}

\begin{figure}[t]
    \centering
    \includegraphics[width=0.95\linewidth]{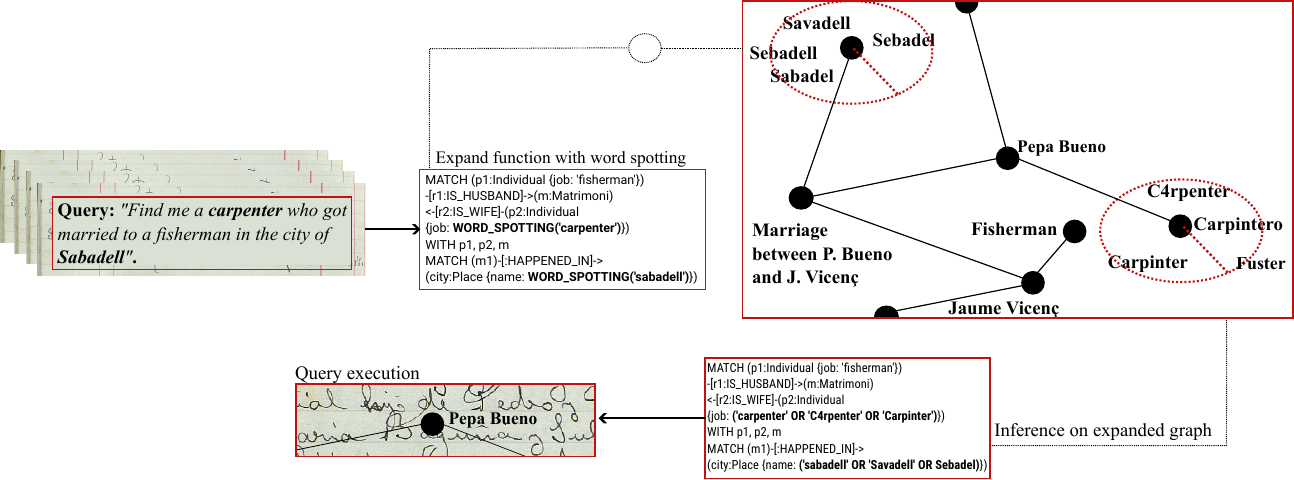}
    \caption{Visual abstract of our contribution. A natural language request is translated into a Cypher\footnotemark{} query. The model handles challenging node types (e.g., names, places) by choosing word-spotting expansion or exact matching. Queries run on the expanded graph, ensuring robust, flexible, and trustworthy retrieval.}
    \label{fig:visual_abstract}
\end{figure}

Historical Document Retrieval is the task that lies at the core of access portals to digital libraries and archives. Either by retrieving high-level features in content-based retrieval \cite{seuret2021icdar}, or low-level features in writer identification \cite{christlein2019icdar} and word spotting \cite{fischer2011transcription}, this task constitutes, for many users, the entry point to valuable historical sources and collections. These systems are central to the construction of historical narratives \cite{epstein2024can,ludolph2016manipulating}, as they offer an inevitably partial view in which generalization and specificity must be balanced. \cite{wang2023survey}.

While many industries have adopted neural-based representations integrated into their search engines \cite{karakurt2025retrieval}, most GLAM institutions (Galleries, Libraries, Archives, and Museums) still rely on ontology and metadata-based approaches for organizing and accessing their information \cite{francearchives_2026,arxius_en_linia_2026,loc_catalog_search_2026}.  We attribute this apparent stagnation to the aforementioned tension: while analyzing historical data requires strong generalization capabilities due to its multi-domain, multi-source nature and wide temporal variance, many of the most interesting and important elements in historical documents are not discoverable through Pattern-Recognition methods. Rather, they are facts expressed through symbolic entities and relations, for which only an ontological approach can guarantee their presence or absence within a database. In recent years, agentic perspectives on language modeling have been shown to serve as a bridge between flexibility and robustness \cite{singh2025agentic}; however, most of their applications are still found in digitally native documents \cite{edge2024local}. The idea of leveraging a knowledge graph for an agent to index is appealing, but unrealistic in the presence of recognition artifacts and errors that can corrupt the inherent structure of such a graph. For an agentic (therefore generalizable), graph-based (therefore grounded) system to operate effectively in historical document analysis, it must account for inaccuracies in handwritten text recognition technologies, which may corrupt its structure and yield numerous false negatives.
\footnotetext{Language designed for indexing graph databases.}

In this work, we propose a seminal implementation of a language model agent that can access historical data by exploiting a historical knowledge graph via code execution. In contrast to common GraphRAG implementations~\cite{singh2025agentic}, we construct the graph from historical handwritten documents; therefore, errors are not occasional corrupted pieces of information but a consistent characteristic of the graph. As shown in Figure~\ref{fig:visual_abstract}, we integrate word-spotting techniques into the agentic process. In this way, we provide the code-generation model with the option to expand certain nodes of the graph to create a meta-node consisting of all possible variations of a given word. We validate this perspective using a combination of different HTR transcriptions, human transcriptions, multiple word-spotting techniques, diverse graph representations, and code generators on the Esposalles \cite{romero2013esposalles} dataset. Accordingly, we summarize our contributions as follows:

\begin{enumerate}
    \item We introduce the first GraphRAG architecture designed explicitly for noisy historical knowledge graphs, where query execution is dynamically augmented through word-spotting–based literal expansion at runtime.
    
    \item We contribute with the human annotation of 100 curated book-level questions on the Esposalles dataset, which serves as use case for our method, consisting of extractive tasks, statistical inference and multi-hop queries.

\end{enumerate}

The rest of the work is organized as follows: in Section~\ref{sec:related} an overview of Retrieval Augmented Generation literature is done, including its potential on managing digital archives. Our method is defined in Section~\ref{sec:method}, where we introduce from the query formulation to the code generation and word spotting injection. In Section~\ref{sec:exp} different data structures that will be employed are presented, and the Esposalles-DocVQA dataset is introduced as a showcase of our method's performance on a real environment. Results and conclusions are to be found in Sections \ref{sec:res} and \ref{sec:conc} respectively.

\section{Related Work}
\label{sec:related}
Pretrained generative models for automatic text generation based on Transformers rely solely on knowledge acquired during pretraining. This makes them prone to generating plausible but incorrect information, a phenomenon known as hallucination \cite{kalai2025language}, and limits their ability to update knowledge dynamically. In contrast, traditional information retrieval systems efficiently provide relevant documents but lack the ability to synthesize or integrate this information into coherent and narrative outputs \cite{gienapp2024evaluating}.

In 2020, Retrieval-Augmented Generation (RAG) systems were introduced \cite{lewis2020retrieval}. These architectures combine an external retrieval process to obtain factual evidence with a generative model capable of incorporating this information into novel and coherent responses \cite{lewis2020retrieval}. This approach aims to ensure factual fidelity in generated content while overcoming the limitations of purely generative or retrieval-based systems.

Early hybrid systems such as DrQA \cite{chen2017reading} combined retrieval with limited generation, often restricted to selecting textual fragments from retrieved documents. Later, REALM \cite{guu2020retrieval} represented a significant advancement by integrating dense retrieval during pretraining with generation, improving the semantic alignment between retrieved information and generated text. This approach was further extended by RAG, which explicitly leverages dense passage retrieval combined with generative \textit{Transformer}-based models such as BART \cite{lewis2020bart}, achieving a more coherent integration between retrieval and generation. Recent trends point to advanced hybrid approaches that combine traditional lexical and dense signals to enhance both the precision and coverage of RAG systems \cite{hsu2025dat}.


Embedding-based methods often struggle with multi-faceted textual queries \cite{wang2024balanced}. Probabilistic approaches improve results \cite{chun2023improved} but still face challenges in reasoning over structured information, complex operations, and transparency in passage selection \cite{borgeaud2022improving,izacard2023atlas}. Agentic workflows address these issues by integrating reasoning and query actions. ReAct alternates step-by-step reasoning with queries to reduce factual errors \cite{yao2023react}, Self-Ask decomposes questions into sub-queries and validates consistency \cite{press2022measuring}, and Chain-of-Agents coordinates specialized LLMs for planning, retrieval, and verification in long contexts \cite{zhang2024chain}. These approaches fall under Agentic RAG, which formalizes planning, reflection, and tool usage in the workflow \cite{singh2025agentic}.
Recent work at the intersection of vision and symbolic reasoning has attempted to bridge high-level semantic understanding with visual perception through program synthesis. Representative approaches include the neuro-symbolic framework VisProg and the code-based reasoning system ViperGPT. Both ViperGPT \cite{suris2023vipergpt} and VisProg \cite{gupta2023visual} provide zero-shot frameworks that answer visual queries by generating Python programs composed of API calls to pretrained perception models, such as object detectors and attribute classifiers. Instead of reasoning visually, these systems define the logic of visual checks and delegate execution, enabling interpretability, compositionality, and strong generalization without additional training.

In parallel, the integration of Knowledge Graphs (KGs) introduces a structured layer that facilitates multi-hop reasoning, allowing inference across multiple chains of relationships rather than only direct connections. GreaseLM demonstrates that combining a language model with a Graph Neural Network enhances entity inference \cite{wu2020comprehensive}. KG\textsuperscript{2}RAG organizes and expands passages using an external KG \cite{zhu2025knowledge}, while GraphRAG generates graphs from free text and retrieves subgraphs as explainable evidence during generation \cite{edge2024local}. Our work is situated within this context, employing a hybrid RAG approach that leverages routines to a graph database to produce reliable insights grounded in factual knowledge.

\section{Method}
\label{sec:method}

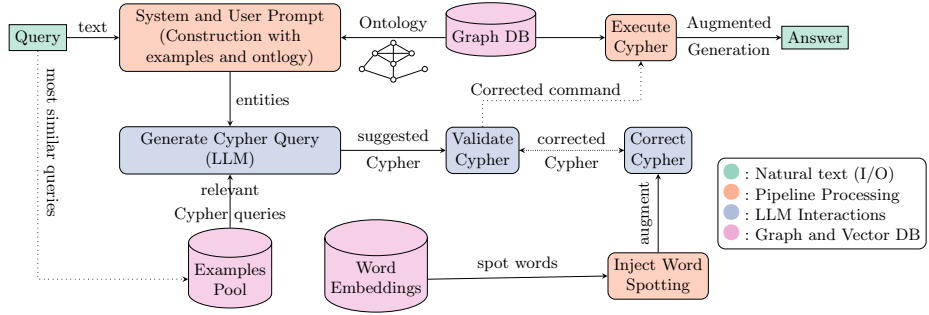
\begin{figure}[t]
    \centering
    \resizebox{\textwidth}{!}{\begin{tikzpicture}[
    >=stealth,
    node distance=2cm
]

\definecolor{colorTextIO}{RGB}{102,194,165} 
\definecolor{colorPipeline}{RGB}{252,141,98} 
\definecolor{colorLLM}{RGB}{141,160,203} 
\definecolor{colorDB}{RGB}{231,138,195} 
\definecolor{colorWordSpotting}{RGB}{255,165,0} 

\node[draw, fill=colorTextIO!40] (query) at (-10,0) {Query};
\node[draw, fill=colorPipeline!40, anchor=west, align=center, text width=4cm, rounded corners] 
    (pipeline_academic) at ($(query.east) + (1,0)$) 
    {System and User Prompt\\(Construction with examples and ontlogy)};
\node[draw, cylinder, shape border rotate=90, anchor=west, aspect=0.24, fill=colorDB!40] 
    (db) at ($(pipeline_academic.east) + (2,0)$) {Graph DB};

\node[draw, rounded corners, align=center, text width=4cm, fill=colorLLM!40] 
    (generate_Cypher) at ($(pipeline_academic.south) + (0,-1.5)$)
    {Generate Cypher Query\\(LLM)};

\node[
    draw,
    cylinder,
    shape border rotate=90,
    aspect=0.35,
    align=center,
    fill=colorDB!40
] 
(examples_pool) at ($(generate_Cypher.south) + (0,-2)$)
{Examples\\Pool};

\node[
    draw,
    cylinder,
    shape border rotate=90,
    aspect=0.35,
    align=center,
    fill=colorDB!40,
    anchor=west
] 
(word_embeddings) at ($(examples_pool.east) + (1, 0)$)
{Word\\Embeddings};

\node[draw, rounded corners, align=center, anchor=west, fill=colorLLM!40]
    (validate_Cypher) at ($(generate_Cypher.east) + (2,0)$)
    {Validate\\Cypher};
\node[draw, rounded corners, align=center, anchor=west, fill=colorLLM!40]
    (correct_Cypher) at ($(validate_Cypher.east) + (2,0)$)
    {Correct\\Cypher};

\node[draw, rounded corners, align=center, fill=colorPipeline!40, anchor=north]
    (inject_word_spotting) at ($(correct_Cypher.south) + (0,-1.5)$)
    {Inject Word\\Spotting};

\node[draw, rounded corners, align=center, fill=colorPipeline!40]
(execute) at ($(db.east) + (2,0)$)
{Execute\\Cypher};

\draw[->] (db.east) -- (execute.west);
\draw[->] (query.east) -- (pipeline_academic.west) node[midway, above] {text};
\draw[->] (pipeline_academic.south) -- (generate_Cypher.north) node[midway, right] {entities};
\draw[->] (examples_pool.north) -- (generate_Cypher.south) 
    node[midway, above] {relevant} node[midway, below] {Cypher queries};
\draw[->] (generate_Cypher.east) -- (validate_Cypher.west) 
    node[midway, above] {suggested} node[midway, below] {Cypher};
\draw[dotted, ->] (validate_Cypher.east) -- (correct_Cypher.west) node[midway, above] {corrected};
\draw[dotted, ->] (correct_Cypher.west) -- (validate_Cypher.east) node[midway, below] {Cypher};
\draw[<-] (pipeline_academic.east) -- (db.west) node[midway, above] {Ontology};
\draw[dotted, ->] (query.south) |- (examples_pool.west)
    node[pos=0.2, sloped, above] {most similar queries};

\draw[->] (word_embeddings.east) -- (inject_word_spotting.west) 
    node[midway, above] {spot words};
\draw[->] (inject_word_spotting.north) -- (correct_Cypher.south) 
    node[midway, above, sloped] {augment};

\coordinate (midpoint) at ($(validate_Cypher.north) + (0,0.5)$);
\draw[dotted, ->] (validate_Cypher.north) -- (midpoint) -| (execute.south) node[pos=0.2, sloped, above] {\footnotesize{Corrected command}};

\node[circle, draw, inner sep=1pt] (g1) at ($(pipeline_academic.east)!0.5!(db.west) + (0,-0.1)$) {};
\node[circle, draw, inner sep=1pt] (g2) at ($(g1) + (-0.35,-0.2)$) {};
\node[circle, draw, inner sep=1pt] (g3) at ($(g1) + (0.35,-0.2)$) {};
\node[circle, draw, inner sep=1pt] (g4) at ($(g1) + (0,-0.4)$) {};
\node[circle, draw, inner sep=1pt] (g5) at ($(g2) + (-0.25,-0.3)$) {};
\node[circle, draw, inner sep=1pt] (g6) at ($(g3) + (0.25,-0.3)$) {};
\node[circle, draw, inner sep=1pt] (g7) at ($(g4) + (0,-0.3)$) {};
\draw (g1) -- (g2); \draw (g1) -- (g3); \draw (g1) -- (g4);
\draw (g2) -- (g3); \draw (g2) -- (g4); \draw (g3) -- (g4);
\draw (g2) -- (g5); \draw (g4) -- (g7); \draw (g5) -- (g7); \draw (g6) -- (g7);

\node[draw, align=center, anchor=west, fill=colorTextIO!40]
    (answer) at ($(execute.east) + (2,0)$)
    {Answer};
\draw[->] (execute.east) -- (answer.west) node[midway, above] {Augmented} node[midway, below] {Generation};

\begin{scope}[shift={($(answer.east |- examples_pool.south) + (-2.5,2)$)}]
\node[draw, align=left, anchor=west, rounded corners, inner sep=3pt] (legend) at (0,0) {
    \tikz{\fill[colorTextIO!70] (0,0) rectangle (0.3,0.3);} : Natural text (I/O)\\
    \tikz{\fill[colorPipeline!70] (0,0) rectangle (0.3,0.3);} : Pipeline Processing\\
    \tikz{\fill[colorLLM!70] (0,0) rectangle (0.3,0.3);} : LLM Interactions\\
    \tikz{\fill[colorDB!70] (0,0) rectangle (0.3,0.3);} : Graph and Vector DB
};
\end{scope}

\end{tikzpicture}}
    \caption{Scheme of our proposed approach, including the word spotting injection step.}
    \label{fig:metode}
\end{figure}

Figure~\ref{fig:metode} overviews the proposed architecture that is inspired by the principles of Agentic GraphRAG architectures, the components of such architecture are: 
\paragraph{\textbf{Graph Construction (GraphDB)}}

The graph construction process consists of injecting nodes and relations extracted from a handwritten historical corpus. This involves leveraging HTR and NER to identify people, places, events, and their interconnections within the texts. The extracted entities are modeled as nodes, while their semantic relationships are represented as edges. These structured elements are then ingested into a graph database, which serves as the backbone for indexing, enabling efficient querying, navigation, and knowledge discovery over the historical data.

\paragraph{\textbf{Ontology and Graph (System and User Prompt)}}
Given any query, the system is provided with a schema that represents the ontology of the graph database being used. This schema, shall contain structured and ontological information about the database. The graph models the domain as a property graph in which nodes represent entities such as individuals, documents, events, and locations, each annotated with properties (e.g., name, year, file\_name...) depending on the historical source. Relationships between nodes are explicitly typed (e.g., family relationships, locations...) and encode the semantics of how entities are connected within the dataset. The schema provided to the system is a human-readable, textual representation of this ontology. It describes node labels, relationship types, properties, and cardinalities, offering a high-level overview of the graph structure that enables the LLM to understand the underlying domain model, ensuring that generated relationship patterns and property usage are consistent with the actual topology of the graph.


\paragraph{\textbf{Generate Cypher Query}}
The system employs a zero-shot in-context learning paradigm to facilitate the translation from natural language questions to syntactically correct Cypher queries. A set of question-query pairs for each graph structure are elaborated, which serve as contextualized examples for the mapping between the user intent expressed in natural language and the corresponding Cypher logic. During query generation, these examples are injected into the prompt context alongside the graph schema and conversation history, allowing the LLM to identify analogous patterns between the input question and the provided examples. This approach enables the model to generate queries that adhere to established conventions for the specific database structure, including proper use of node labels, relationship directions, property filters, and result limiting clauses.

\paragraph{\textbf{Validation Cypher}}

The validation pipeline applies a multi-stage verification process with iterative error correction before query execution. Generated Cypher queries are first checked syntactically by submitting them to the graph database planner without executing them, allowing the detection of malformed clauses or invalid constructs. They are then validated against the graph schema to ensure that relationship patterns conform to the predefined topology, automatically correcting relationship directions when necessary.
The validation mechanism operates within a correction cycle controlled by an iteration counter. When validation fails, the system sends the query to an error correction phase where the LLM is prompted to fix the identified issues while preserving the query's semantic intent. The corrected query then re-enters validation for verification. This cycle continues until either validation succeeds without errors or a maximum iteration threshold is exceeded, preventing infinite correction loops. 

\paragraph{\textbf{Inject Word Spotting}}

The Word Spotting module implements a query expansion mechanism that enriches generated Cypher queries by matching string literals against a glossary of textual variants obtained through word spotting. For each transcribed word, orthographic variants are clustered using an embedding-based similarity threshold, addressing a central challenge in historical document retrieval: OCR errors, spelling variability, and transcription inconsistencies that prevent exact string matching from retrieving relevant records. The expansion process is parameterized by an embedding specification that determines both the similarity metric (character-level or semantic) and the reference corpus (ground truth or OCR-derived text).

During query generation, the LLM marks literals that require approximate matching. At execution time, the system computes embeddings for each marked literal using either PHOC for character-level similarity or MPNet for semantic similarity, and retrieves similar variants from a vector database above a configurable threshold. The original Cypher query is then refactored by replacing marked literals with case-insensitive regular expressions covering all retrieved variants, while preserving the logical structure of the query. For complex inline patterns, matching conditions are extracted into dedicated \texttt{WHERE} clauses to ensure syntactic correctness. We formalize the word-spotting expansion operator as follows.

\paragraph{\textbf{Formalization of the Expansion Operator.}}
Let $q$ denote the original Cypher query generated by the LLM.
Let $L = \{l_1, \dots, l_n\}$ be the set of string literals in $q$
explicitly marked for expansion. For each literal $l \in L$,
the word-spotting operator $\mathrm{WS}(\cdot)$ returns a set
of textual variants:

\[
\mathrm{WS}(l) \rightarrow \{v_1, \dots, v_k\}.
\]

We define the expanded query $q'$ as the transformation of $q$
in which every marked literal $l$ is replaced by a regular
expression matching the disjunction of its variants:

\[
l \;\mapsto\; \texttt{regex}(v_1 \lor \dots \lor v_k).
\]

Thus, $q'$ preserves the original logical structure of $q$
while relaxing exact string constraints through embedding-based
variant expansion.
\paragraph{\textbf{Execute Cypher}}
Query execution represents the final phase where the validated and potentially enriched Cypher statement is submitted to the graph database for actual data retrieval. The retrieved database records are fed alongside the original user question and the conversation history into a final LLM invocation that synthesizes a user-friendly response. The response generation is designed to present results concisely when documents are found, or provide direct, clear explanations when no relevant information exists, maintaining conversational context throughout multi-turn interactions.

\section{Experimental Setup: A Use Case on the Esposalles Database}
\label{sec:exp}
In this work, we make use of the Esposalles database \cite{romero2013esposalles} to validate the performance of our system in a real-world environment. Rather than using a standard Named Entity Recognition baseline, we will be employing a data set-up organized as follows.

\subsection{The Esposalles Graph Database}
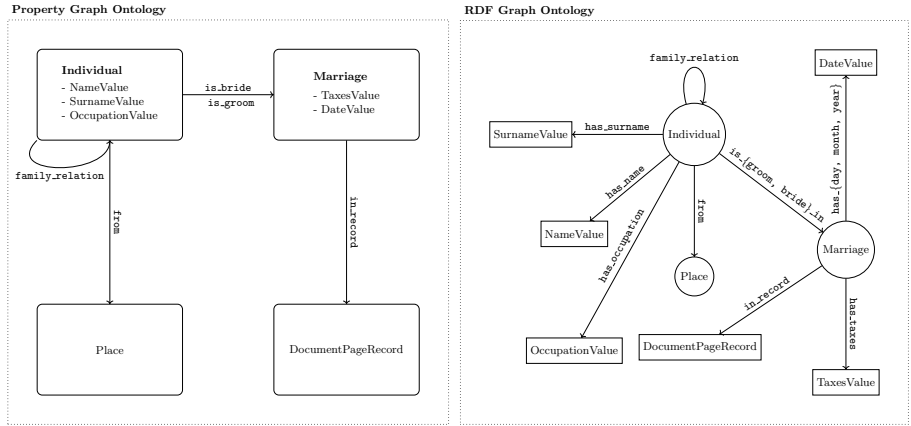
\begin{figure}[t]
    \centering
    \begin{tabular}{cc}
        \resizebox{0.49\textwidth}{!}{\begin{tikzpicture}[node distance=6cm,
                    nodebox/.style={rectangle, rounded corners, draw, minimum width=40mm, minimum height=25mm, align=left, inner sep=5pt, font=\small},
                    prop/.style={rectangle, draw, minimum width=30mm, minimum height=8mm, align=center, font=\small}]
    
    \def\marriagedist{2.5cm}  
    \def\recorddist{4.5cm}      
    
    \node[nodebox] (Individual) {
        \textbf{Individual} \\[2mm]
        \begin{tabular}{@{}l@{}}
            - NameValue \\
            - SurnameValue \\
            - OccupationValue \\
        \end{tabular}
    };
    
    \node[nodebox] (Marriage) [right=\marriagedist of Individual] {
        \textbf{Marriage} \\[2mm]
        \begin{tabular}{@{}l@{}}
            - TaxesValue \\
            - DateValue
        \end{tabular}
    };
    
    \node[nodebox] (DocumentPageRecord) [below=\recorddist of Marriage] {DocumentPageRecord};
    \node[nodebox] (placeNode) [below=\recorddist of Individual] {Place};
    \draw[->] 
        (Individual.east) -- node[midway, above]{\footnotesize\texttt{is\_bride}}
                              node[midway, below]{\footnotesize\texttt{ is\_groom}} 
        (Marriage.west);
    
    \draw[->] (Individual.south west) .. controls ++(-1,-1) and ++(0,-1) .. (Individual.south) 
        node[pos=0.5, below]{\footnotesize\texttt {family\_relation}};
    
    \draw[->] (Marriage.south) -- (DocumentPageRecord.north) node[midway, above, sloped]{\footnotesize\texttt{in\_record}};
    \draw[->] (Individual.south) -- (placeNode.north) node[midway, above, sloped]{\footnotesize\texttt{from}};
    \node[draw, dotted, inner sep=8mm, fit=(Individual) (Marriage) (DocumentPageRecord)] (box) {};
    \node[anchor=south west] at (box.north west) {\textbf{Property Graph Ontology}};
\end{tikzpicture}}  &  
        \resizebox{0.49\textwidth}{!}{\begin{tikzpicture}[node distance=2.5cm, 
                    entity/.style={circle, draw, minimum size=7mm},
                    value/.style={rectangle, draw, minimum size=7mm},
                    every label/.style={font=\footnotesize}]
    \node[entity] (Individual) {Individual};
    \node[entity] (Marriage) [below right=2cm and 3cm of Individual] {Marriage};
    \node[entity] (Place) [below=of Individual] {Place};

    \node[value] (NameValue) [below left=of Individual] {NameValue};
    \node[value] (SurnameValue) [left=of Individual] {SurnameValue};
    \node[value] (OccupationValue) [below=of NameValue] {OccupationValue};
    \node[value] (TaxesValue) [below=of Marriage] {TaxesValue};
    \node[value] (DateValue) [above=4cm of Marriage] {DateValue};
    \node[value] (DocumentPageRecord) [below left=of Marriage] {DocumentPageRecord};

    \draw[->, loop above] (Individual) to node[above]{\footnotesize\texttt{family\_relation}} (Individual);
    \draw[->] (Individual) -- (Marriage) node[midway, above, sloped]{\footnotesize\texttt{is\_\{groom, bride\}\_in}};
    \draw[->] (Individual) -- (Place) node[midway, above, sloped]{\footnotesize\texttt{from}};
    \draw[->] (Individual) -- (NameValue) node[midway, above, sloped]{\footnotesize\texttt{has\_name}};
    \draw[->] (Individual) -- (SurnameValue) node[midway, above, sloped]{\footnotesize\texttt{has\_surname}};
    \draw[->] (Individual) -- (OccupationValue) node[midway, above, sloped]{\footnotesize\texttt{has\_occupation}};
    \draw[->] (Marriage) -- (TaxesValue) node[midway, above, sloped]{\footnotesize\texttt{has\_taxes}};
    \draw[->] (Marriage) -- (DateValue) node[midway, above, sloped]{\footnotesize\texttt{has\_\{day, month, year\}}};
    \draw[->] (Marriage) -- (DocumentPageRecord) node[midway, above, sloped]{\footnotesize\texttt{in\_record}};

    \node[draw, dotted, inner sep=8mm, fit=(Individual) (Marriage) (Place) 
          (NameValue) (SurnameValue) (OccupationValue) (TaxesValue) (DateValue) (DocumentPageRecord)] (box) {};

    \node[anchor=south west] at (box.north west) {\textbf{RDF Graph Ontology}};
\end{tikzpicture}} 
    \end{tabular}
\caption{Property graphs simplify Cypher queries, while RDF enables more complex searches via structured relationships.}
    \label{fig:ontologies}
\end{figure}

Two approaches of graph modeling have been defined in this research. On one hand, in property-based graphs (see Figure~\ref{fig:ontologies}, left)  names and surnames are stored directly as properties of the individual nodes, resulting in a more compact graph structure with fewer nodes and relationships. On the other hand, Resource-Description Framework (see Figure~\ref{fig:ontologies}, right) represents each individual by a unique identifier, while given names and surnames are modeled as distinct value nodes. Individuals are connected to these name nodes through explicit relationships, allowing names to be queried independently.

\begin{table}[t]
\centering

\setlength{\tabcolsep}{4pt} 

\begin{tabular}{ccc} 
\begin{subtable}[c]{0.32\textwidth}
\centering
\caption{HTR accuracy}
\label{tab:htr_accuracy}
\begin{tabular}{l r}
\toprule
\textbf{Name} & \textbf{Acc. (\%)} \\
\midrule
EsposallesGT   & 100.0 \\
HTR4           & 98.0  \\
HTR2           & 95.7  \\
HTR1           & 78.3  \\
HTR0           & 69.6  \\
\bottomrule
\end{tabular}
\label{tab:transcriptions}
\end{subtable} &

\begin{subtable}[c]{0.34\textwidth}
\centering
\caption{Word embeddings}
\label{tab:word_embeddings}
\begin{tabular}{l l}
\toprule
\textbf{Embedding} & \textbf{Transcr.} \\
\midrule
MPNET & GT   \\
MPNET & HTR4 \\
MPNET & HTR2 \\
MPNET & HTR1 \\
MPNET & HTR0 \\
PHOC               & GT   \\
PHOC               & HTR4 \\
PHOC               & HTR2 \\
PHOC               & HTR1 \\
PHOC               & HTR0 \\
\bottomrule
\end{tabular}
\label{tab:embeddings}
\end{subtable} &

\begin{subtable}[c]{0.30\textwidth}
\centering
\caption{Graph databases}
\label{tab:graph_databases}
\begin{tabular}{l c}
\toprule
\textbf{Name} & \textbf{Attr.} \\
\midrule
Property-GT & \xmark \\
RDF-GT   & \cmark \\
RDF-HTR4 & \cmark \\
RDF-HTR2 & \cmark \\
RDF-HTR1 & \cmark \\
RDF-HTR0 & \cmark \\
\bottomrule
\end{tabular}
\label{tab:rdf}

\end{subtable}

\end{tabular}
\caption{Overview of transcription accuracy, word embeddings, and graph configurations}

\label{tab:overview_tables}
\end{table}

\textbf{Esposalles Handwritten Recognition} -- We employ multiple Handwritten Text Recognition (HTR) models to generate noisy versions of the same database. This approach allows us to evaluate the robustness of our method against transcription errors and HTR inaccuracies. Specifically, we utilize off-the-shelf ViT-based HTR models~\cite{rodriguez2025ocr} (0, 1, 2 and 4) trained to achieve different levels of accuracy on the Esposalles database, as summarized in Table~\ref{tab:transcriptions}.

\textbf{Word Embeddings} -- For word spotting, we generate word embeddings from both clean and noisy transcriptions. Table~\ref{tab:embeddings} illustrates how we leverage Pyramid Histogram of Characters (PHOC) embeddings~\cite{almazan2014word} and a semantic search word model, MPNET~\cite{song2020mpnet}, across HTR outputs with varying accuracies. This setup enables us to assess the robustness of different word spotting methods to transcription noise and to evaluate the relative advantages of each embedding approach in various scenarios.

\textbf{Graph Databases} -- For each HTR transcription, we construct a corresponding graph database. In this work, we evaluate the document processing pipeline under the assumption of human-assisted graph construction, a common practice in many digital archives that already maintain interconnected sets of curated labels and entity annotations. Accordingly, we rely on ground-truth named entity annotations rather than performing automatic entity extraction. Our focus is query robustness under transcription noise, therefore automatic graph construction remains orthogonal and is left for future work.  Table~\ref{tab:rdf} details how we build property-based graphs using ground-truth transcriptions, representing perfect HTR accuracy. We also construct Resource Description Framework (RDF) graphs for HTR transcriptions at different accuracy levels, including ground truth, to assess whether the advantages of RDF graphs stem solely from perfect transcriptions or whether the approach remains robust under noisy conditions.


\subsection{The Esposalles DocVQA Evaluation Dataset}
\label{sec:esp_dataset}
A contribution of this work is the creation of an evaluation dataset comprising 100 manually curated question–answer pairs derived from the Esposalles records. The dataset is designed to evaluate the capabilities of the RAG systems proposed in this paper.

Questions were designed to reflect realistic analytical and lookup needs when working with structured archival records. Instead of focusing exclusively on direct fact retrieval, the dataset was intentionally designed to include questions that require different levels of aggregation, relational navigation, constraint filtering, and multi-step inference across records. This approach ensures coverage of both simple and operationally complex information needs that arise when interacting with historical registries.

All questions were strictly constrained to the information present in the Esposalles records and were validated to be answerable without external knowledge. Each reference answer was manually derived and verified against the source records to ensure correctness and reproducibility. This controlled construction process ensures that the evaluation measures how effectively each system retrieves relevant information and derives answers from the indexed material.

While traditional vector-based RAG systems perform well on semantic similarity tasks and direct entity lookups, they are not designed to answer reliably queries that require structured operations over multiple records or fields. The dataset was therefore designed to expose these differences in capability across retrieval and reasoning strategies.
The dataset includes four main question types designed to test different retrieval and reasoning skills. Statistical aggregation questions (46\%) involve counting, ranking, or analyzing distributions across records. Entity and relationship queries (31\%) target specific entities, their attributes, and relationships, including temporal, spousal, and parental connections. Multi-hop reasoning questions (11\%) require inference across multiple relationships, such as tracking inherited professions or generational patterns. Complex filtered queries (12\%) combine multiple constraints, pattern matching, or range-based filters with aggregation, challenging the system to integrate multiple criteria for accurate retrieval.

\section{Results}
\label{sec:res}
\subsection{Metrics and evaluation}

To assess the quality of generated answers across different RAG configurations, we employed a dual evaluation strategy combining deterministic similarity metrics with LLM-based semantic judgement. This approach balances the objectivity and reproducibility of automated metrics with the nuanced understanding of language models capable of handling linguistic variations inherent in historical documents.

\subsubsection{Deterministic Entity-Based Evaluation}
The first evaluation metric is our proposed recall-oriented entity matching algorithm for historical document question-answering, termed NERO-ANLS (Named Entity – Recall-Oriented Average Normalized Levenshtein Similarity). It extends the original ANLS metric~\cite{mathew2021docvqa}, which is unsuited for verbose generative models due to its sensitivity to extraneous text, by focusing on structured entity extraction and comparison. The deterministic procedure leverages the \texttt{\seqsplit{ca\_core\_news\_trf}} named entity recognizer \cite{armengol2021multilingua} to evaluate answer quality through entity-level similarity.
\paragraph{\textbf{Entity Extraction Pipeline}}
Evaluation begins by extracting key elements from both ground truth and generated answers, including numerical entities (dates, quantities, counts), named entities such as person names (PER), locations (LOC), organizations (ORG), and temporal expressions (DATE), as well as key lexical items like proper nouns, meaningful nouns (excluding stopwords), and profession terms. Text preprocessing includes markdown removal, normalization of whitespace, and case-folding. For expected answers (typically concise given the DocVQA dataset), we extract all meaningful words exceeding 2 characters to ensure proper coverage. For generated answers (often extensive and verbose), we focus on named entities and key nouns to identify relevant information.



\begin{figure}[t]
    \centering
    \begin{subfigure}[b]{0.49\linewidth}
        \centering
        \includegraphics[width=\linewidth]{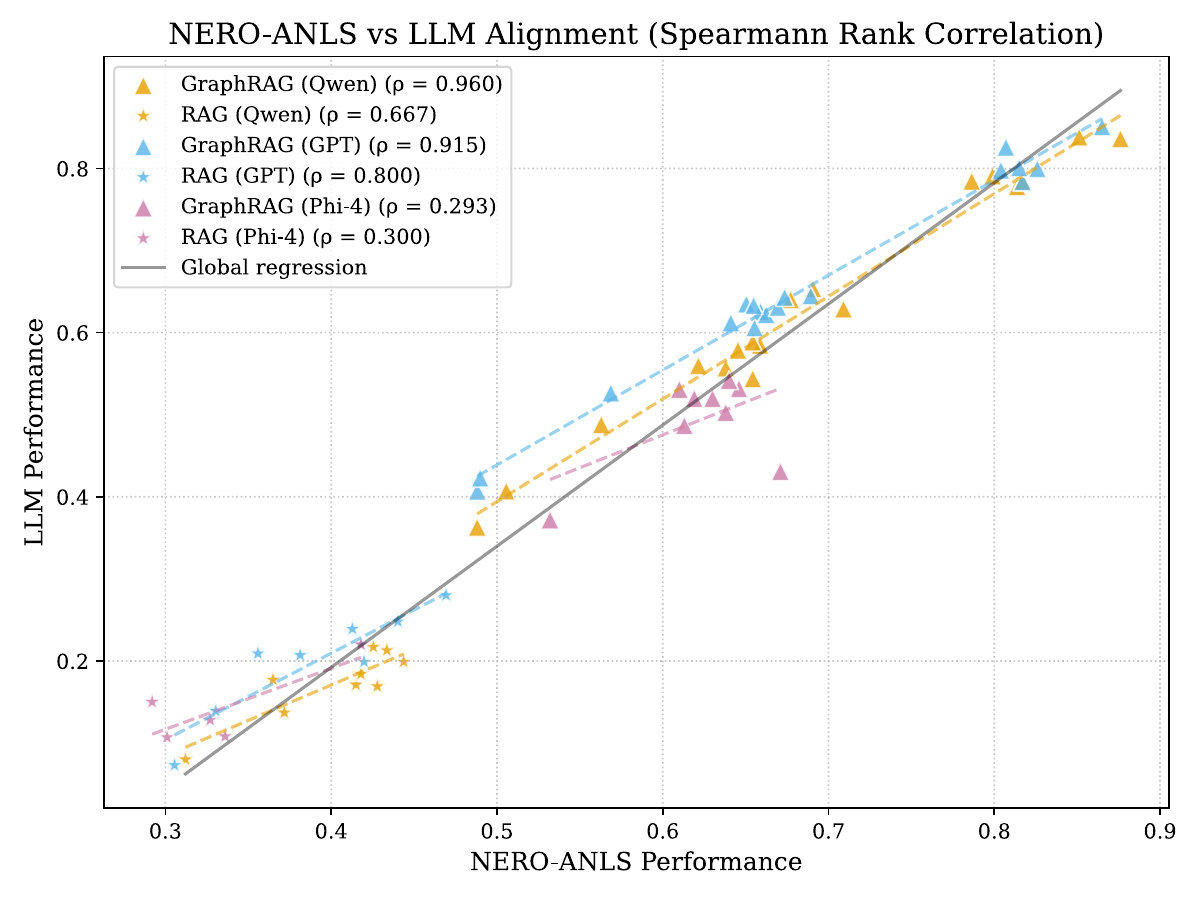}
        \caption{Alignment with LLM-as-a-Judge across HTR0–HTR4 and RDF/Property graphs.}
        \label{fig:subfig2}
    \end{subfigure}
    \begin{subfigure}[b]{0.49\linewidth}
        \centering
        \includegraphics[width=\linewidth]{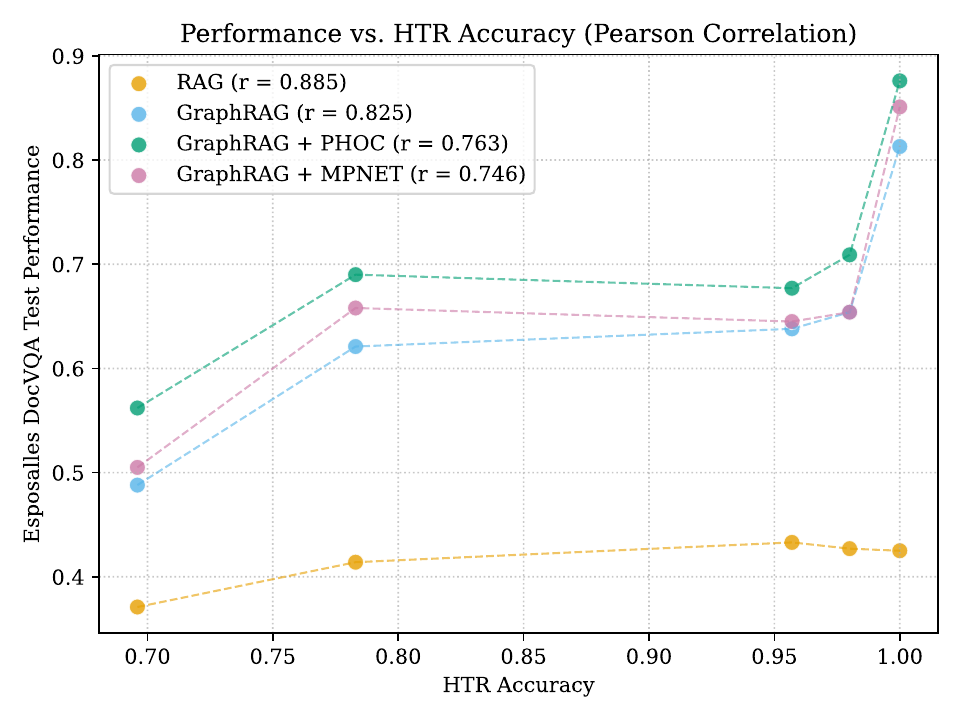}
        \caption{RAG, GraphRAG, and GraphRAG+Word Spotting performance under varying HTR accuracies.}
        \label{fig:subfig1}
    \end{subfigure}
    \hfill

    \caption{LLM-as-a-Judge alignment and system performance vs. HTR accuracy.}
    \label{fig:performance_htr}
\end{figure}
\paragraph{\textbf{Similarity Computation}}
Similarity is computed in two stages. First, exact matching finds intersections between normalized expected and generated elements. Second, unmatched elements are compared using Levenshtein-distance-based fuzzy matching with a 0.6 threshold, accommodating common orthographic variations in historical transcriptions (e.g., ``Pagès'' vs. ``pages'', ``Terrassa'' vs. ``jerrassa'').
\paragraph{\textbf{Score Calculation}}
The final score prioritizes recall (coverage of expected information) over precision, reflecting the evaluation goal of assessing whether the system retrieved all necessary information rather than penalizing verbosity:
\begin{equation}
\text{Score} = \begin{cases}
0.95 + 0.05 \times \text{recall} & \text{if recall} \geq 0.95 \\
0.85 + 0.1 \times \text{recall} & \text{if } 0.95 > \text{recall} \geq 0.85 \\
F_\beta & \text{otherwise}
\end{cases}
\end{equation}
where $F_\beta$ is the F-score with $\beta = 3$, where recall is weighted 3 times higher than precision, calculated as:
\begin{equation}
F_\beta = \frac{(1 + \beta^2) \cdot \text{precision} \cdot \text{recall}}{\beta^2 \cdot \text{precision} + \text{recall}}
\end{equation}
and precision is capped at a ratio of 1:5 (expected:generated elements) to avoid penalizing naturally verbose LLM responses. This design acknowledges that historical QA systems should prioritize information completeness over conciseness.

\subsubsection{LLM-as-a-Judge Evaluation}
To complement the deterministic metric and capture semantic equivalences beyond lexical matching, we implemented an LLM-as-a-judge evaluation framework. This approach leverages an understanding of semantic similarity language from the model, context, and domain-specific knowledge to provide human-like quality assessments.

\paragraph{\textbf{Evaluation Protocol and Scoring}}
The LLM judge evaluates system answers against the ground truth using a structured prompt with the user question, reference, and generated answer. Assessment considers entity accuracy (names, places, numbers, dates, professions), completeness, correctness of counts, and minor orthographic variations. Scores range 0–10, with low for missing/incorrect entities, mid for partial correctness, and high for mostly correct answers; maximum scores require all key entities, minor spelling tolerance, and fully accurate counts, prioritizing coverage over style.

\begin{table}[t]
\centering

\resizebox{\textwidth}{!}{
\begin{tabular}{|l|c|l|c|c|c|c|c|c|}
\hline
\multicolumn{3}{|c|}{} & \multicolumn{2}{c|}{\textbf{Qwen3-VL (32B)}} & \multicolumn{2}{c|}{\textbf{GPT-5.1}} & \multicolumn{2}{c|}{\textbf{Phi-4 (15B)}} \\
\cline{4-9}
\multicolumn{3}{|c|}{} & \textbf{NERO-ANLS} & \textbf{LLM} & \textbf{NERO-ANLS} & \textbf{LLM} & \textbf{NERO-ANLS} & \textbf{LLM} \\
\hline
\hline
\multirow{8}{*}{\textbf{RAG}} & \multicolumn{2}{l|}{GT Transcription} & .435 & .217 & {.469} & {.280} & .418 & .220 \\
\cline{2-9}
& \multicolumn{2}{l|}{HTR0 Transcription} & {.371} & { .137 }& .330 & .139 & .336 & .108\\
\cline{2-9}
& \multicolumn{2}{l|}{HTR1 Transcription} & .414 & .171 & .412 & .239 & .292 & .150 \\
\cline{2-9}
& \multicolumn{2}{l|}{HTR2 Transcription} & .433 & .213 & .381 & .207 & .301 & .107 \\
\cline{2-9}
& \multicolumn{2}{l|}{HTR4 Transcription} & .427 & .169 & .419 & .199 & .327 & .128 \\
\hline
\hline
\multirow{9}{*}{\textbf{Graph RAG}} & \multirow{3}{*}{{-}} & Property-GT & .799 & .791 & .807 & .826 & .646 & \textbf{.532} \\
\cline{3-9}
& & RDF-GT & .813 & .778 & \textbf{.865 }& \textbf{.851} &  .638 & .503\\
\cline{3-9}
& & RDF-HTR1 & .621 & .560 & .659 & .626 &.532& .372\\
\cline{2-9}
& \multirow{2}{*}{{PHOC}} & Property-GT & .786 & .784 & .804 & .798 & .613 &.531 \\
\cline{3-9}
& & RDF-GT & \textbf{.876} & .836 & .817 & .784 & .646 & .542\\
\cline{3-9}
& \multirow{4}{*}{{MPNET}} & RDF-HTR1 & .690 & .653 & .669 & .631 & .631 & .520 \\
\cline{2-9}
& & Property-GT & .799 & .817 & .815 & .801 & .619 & .520\\
\cline{3-9}
& & RDF-GT & .813 & \textbf{.851 }& .826 & .800 & .613  &.487 \\
\cline{3-9}
& & RDF-HTR1 & .690 & .658 & .673 & .643 & \textbf{.671} & .431\\
\hline
\end{tabular}}
\caption{Performance metrics by system (rows) and back-end LLM (columns) for both the deterministic and the LLM-as-a-Judge metric. Bold font means \textbf{best performing method}.}
\label{tab:metrics}
\end{table}



This dual evaluation framework is empirically validated by the strong correlation observed in Figure~\ref{fig:subfig2}. By combining deterministic quantitative scoring with qualitative LLM-based assessment, it provides a robust and reproducible evaluation of RAG system performance across diverse question types and varying levels of historical document complexity.

\begin{figure}[t]
    \centering
    \includegraphics[width=0.99\linewidth]{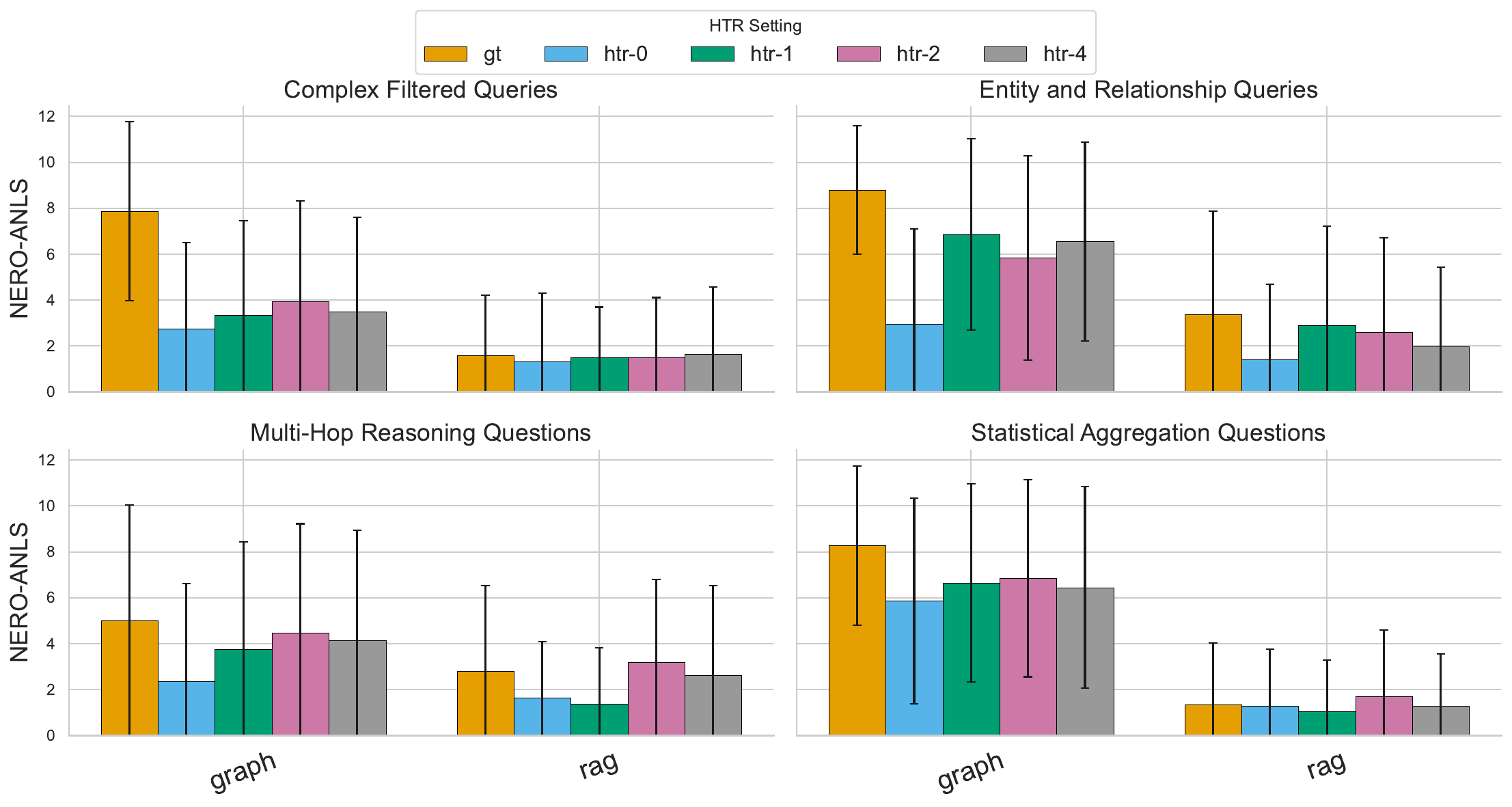}
    \caption{Barplot for GraphRAG and RAG comparison on each question type and HTR transcription.}
    \label{fig:compare_barplot}
\end{figure}

\subsection{Quantitative Analysis}
Once the experimental setup is defined, the global results are summarized in Table~\ref{tab:metrics}. First, we observe a strong alignment between the deterministic (NERO-ANLS) and the LLM-based evaluation protocols. Specifically, by computing the Spearman correlation coefficient, we obtain high correlation values (see Figure~\ref{fig:subfig2}), indicating that both evaluation protocols are well-suited for assessing the capabilities of RAG and GraphRAG systems on extractive and reasoning-based tasks, with some alignment issues when analyzing the lowest-performing models (Phi-4). Second, we note a clear performance ceiling for RAG systems, which struggle to surpass the $0.50$ threshold in terms of NERO-ANLS. This observation aligns closely with the characteristics of the Esposalles DocVQA Evaluation dataset introduced in Section~\ref{sec:esp_dataset}, where $\sim$30\% of the question–answer pairs correspond to entity and relationship queries. As will be further discussed in the qualitative analysis, this suggests that the performance of RAG systems is largely constrained to extractive, self-contained, and well-grounded tasks. Despite this limitation, RAG approaches exhibit significantly higher stability than the proposed GraphRAG method. This behavior can be interpreted either as a greater resilience of RAG systems to degradations in Handwritten Text Recognition quality or as an indication that GraphRAG systems are better positioned to exploit improvements in transcription accuracy. As shown in Figure~\ref{fig:subfig1}, although GraphRAG presents a steeper performance slope, the degradation of RAG performance remains relatively moderate as transcription noise increases.
\begin{table}[t]
\centering
\caption{Comparison of query results without and with PHOC expansion}
\label{tab:phoc_example}
\small
\resizebox{0.85\textwidth}{!}{

\begin{tabular}{p{0.45\textwidth}p{0.45\textwidth}}
\toprule
\textbf{Without PHOC Expansion} & \textbf{With PHOC Expansion} \\
\midrule
\multicolumn{2}{l}{\textit{Question:}} \\
\multicolumn{2}{l}{How many individuals lived in Girona?} \\
\midrule
\multicolumn{2}{l}{\textit{Expected Answer:}} \\
\multicolumn{2}{l}{9 individuals.} \\

\midrule
\multicolumn{2}{l}{\textit{Cypher Query:}} \\
\texttt{\footnotesize MATCH (i:Individual)} & \texttt{\footnotesize MATCH (i:Individual)} \\
\texttt{\footnotesize -[:LIVES]->(l:Place \{name: "girona"\})} & \texttt{\footnotesize -[:LIVES]->(l:Place) WHERE l.name =\textasciitilde} \\
\texttt{\footnotesize RETURN COUNT(i)} & \texttt{\footnotesize '(?i)\textasciicircum(?:girona|gerona|} \\
 & \texttt{\footnotesize gerona|gran|hieronma)\$'} \\
 & \texttt{\footnotesize RETURN COUNT(i)} \\
\midrule
\multicolumn{2}{l}{\textit{Database Records:}} \\
\texttt{\{\textquotesingle population\textquotesingle: 5\}} & \texttt{\{\textquotesingle population\textquotesingle: 9\}} \\
\midrule
\multicolumn{2}{l}{\textit{Generated Answer:}} \\
According to the results found, & Based on the information found, \\
\textbf{5 individuals} lived in Girona. & the number of individuals who lived \\
 & in Girona is \textbf{9}. \\
 \midrule
\multicolumn{2}{p{0.9\textwidth}}{\textit{Exact matching for ``girona'' misses variant spellings, while PHOC expansion generates a regex capturing all mis-transcriptions.}} \\

\midrule

\bottomrule
\end{tabular}
}
\label{tab:quali1}
\end{table}

Contrary to our initial hypothesis, word spotting does not stabilize performance by making HTR accuracy irrelevant. Instead, it consistently improves results across all systems, with a special impact on smaller models, with a large gap for Phi-4 word spotting (from 0.53 to 0.631 in NERO-ANLS when using noisy transcriptions from HTR1). PHOC-based representations are particularly effective for word spotting when compared to MPNet embeddings. Although MPNet is trained for semantic search, it still contributes to performance gains under lower HTR accuracy conditions. We attribute the overall improvement observed in PHOC-augmented GraphRAG systems to the intrinsic diversity and complexity of the Esposalles dataset. As is common in historical corpora, annotation inconsistencies arise due to high variability, contextual ambiguity, and subjective decisions made by annotators.

Within the same results table, we observe a consistent preference for our Word-Spotting–Augmented GraphRAG approach when using Resource Description Framework (RDF) graph modeling. Specifically, RDF achieves superior performance in 7 out of 9 NERO-ANLS comparisons (three LLMs indexed with three word-spotting alternatives in Table~\ref{tab:metrics}), with the only exceptions occurring when using Phi-4. Although property-based schemas offer a simpler structure that can facilitate the generation of syntactically correct queries, RDF-based representations provide greater expressiveness and are better suited to tasks that involve complex relational structures and long-range dependencies within the database. In this sense, the increased modeling complexity does not hinder performance; rather, it constitutes a valuable source of structured information that the model is able to exploit effectively.

With regard to the back-end LLM, which is responsible for both code generation and final answer production, we observe a substantial performance gap between \texttt{Phi-4} and \texttt{Qwen-3}. In contrast, the difference between \texttt{Qwen-3} and \texttt{GPT-5.1} is negligible. This suggests that code inference is, as expected, sensitive to model scale, particularly when moving from smaller to mid-sized models. However, performance appears to plateau beyond approximately 32B parameters, as no measurable improvement is observed on the evaluation dataset when further increasing model size. Additionally, in Figure~\ref{fig:compare_barplot}, the results are disaggregated by question type and the HTR transcriptions used. As expected, Entity and Relationship Queries are the category in which RAG-based systems perform best. The largest performance gap, however, is observed in Statistical Aggregation Questions, which require the ability to analyze the collection holistically (a capability enabled by the graph-based nature of our method) and to perform mathematical operations, which is facilitated by the code-execution component of our approach. With respect to sensitivity to HTR accuracy, we observe a substantial performance gap between ground-truth and noisy transcriptions when addressing complex filtering questions. In contrast, multi-hop reasoning and statistical aggregation questions exhibit remarkable stability across varying transcription quality levels.

\subsection{Qualitative Analysis}
From a qualitative perspective, we present three illustrative examples demonstrating how our approach can improve the indexing and accessibility of digital libraries. First, in Table~\ref{tab:quali1}, we show how common agentic GraphRAG systems may distort historical records due to overreliance on their code inference step. In contrast, our method incorporates a word spotting stage, which in this example transforms the word ``Girona'' into a set of variations detected throughout the document, yielding a more accurate response. Second, in Table~\ref{tab:quali3}, we demonstrate that RAG-based systems still have potential to handle logical queries to some extent. However, a fundamental limitation of RAG is its design constraint: it typically contains only a subset of the documents, which restricts holistic analysis. Achieving full-context reasoning with RAG would require a large context window and significant computational resources. In contrast, graph-based systems, whether combined with code execution or not, enable holistic analysis of document collections more efficiently and comprehensively.
\begin{table}

\centering
\caption{Comparison of query results between GraphRAG and RAG approaches}
\label{tab:quali3}

\label{tab:graphrag_vs_rag}
\small
\resizebox{0.85\textwidth}{!}{

\begin{tabular}{p{0.45\textwidth}p{0.45\textwidth}}
\toprule
\textbf{RAG Approach} & \textbf{GraphRAG Approach} \\
\midrule
\multicolumn{2}{l}{\textit{Question:}} \\
\multicolumn{2}{l}{Where do individuals with the surname Colomer live?} \\
\midrule
\multicolumn{2}{l}{\textit{Expected Answer:}} \\
\multicolumn{2}{l}{Barcelona, Arenys, Pierola, and Viladecans.} \\

\midrule
\multicolumn{2}{l}{\textit{Query/Method:}} \\
Retrieved $N$ document fragments & \texttt{\footnotesize MATCH (p:Individual)} \\
mentioning Colomer & \texttt{\footnotesize -[:SURNAME1]->(surname:value)} \\
(limited sample) & \texttt{\footnotesize MATCH (p)-[:LIVES]->(l:Place)} \\
 & \texttt{\footnotesize WHERE surname.name =\textasciitilde} \\
 & \texttt{\footnotesize '(?i)\textasciicircum(?:colomer|...)\$'} \\
 & \texttt{\footnotesize RETURN l.name, COUNT(p)} \\
\midrule
\multicolumn{2}{l}{\textit{Generated Answer:}} \\
Individuals with the surname Colomer, & Mainly in \textbf{Barcelona},\\
 live in \textbf{Bara}. Specifically, &  where there are  the most records (5), \\
Hieronym Colomer, a tailor from &  \textbf{Arenys}, \textbf{Pierola} and \textbf{Viladecans}.\\
Ribes, lives in Bara. & \\

\midrule
\multicolumn{2}{p{0.9\textwidth}}{\textit{RAG retrieves only a limited set of documents, producing partial answers, while GraphRAG queries all nodes for full coverage.}} \\

\midrule
\bottomrule
\end{tabular}}
\end{table}

\section{Conclusions}
\label{sec:conc}
We presented a framework integrating agentic paradigms and word spotting into Historical Document Analysis, bridging the gap between digitally native and historical documents in graph-based systems. Experiments on the Esposalles DocVQA dataset demonstrate that word spotting improves robustness to noisy transcriptions, particularly when combined with RDF graphs, while system performance stabilizes once sufficient model capabilities are reached.  

Our evaluation shows strong alignment between quantitative metrics and LLM-based assessment, validating both accuracy and interpretability. Even with ground-truth transcriptions, PHOC-based word spotting remains beneficial due to spelling variability and inconsistent annotations. By consolidating orthographic variants within graph nodes, the approach ensures traceable, auditable retrieval, providing holistic and trustworthy access to historical information. These results suggest that agentic graph-based workflows are a promising direction for reliable historical document analysis.

\section*{Acknowledgements}
\small
This work has been partially supported by the Spanish project PID2024-157778OB-I00, Ministerio de Ciencia e Innovación, the Departament de Cultura of the Generalitat de Catalunya, and the CERCA Program / Generalitat de Catalunya. Adrià Molina is funded with the PRE2022-101575 grant provided by MCIN / AEI / 10.13039 / 501100011033 and by the European Social Fund (FSE+). 

\bibliographystyle{splncs04}
\bibliography{mybibliography}

\end{document}